\documentclass[conference]{IEEEtran}
\IEEEoverridecommandlockouts
\usepackage{cite}
\usepackage{amsmath,amssymb,amsfonts}
\usepackage{algorithmic}
\usepackage{graphicx}
\usepackage{textcomp,booktabs,array,makecell}
\usepackage{xcolor}
\def\BibTeX{{\rm B\kern-.05em{\sc i\kern-.025em b}\kern-.08em
    T\kern-.1667em\lower.7ex\hbox{E}\kern-.125emX}}
\begin{document}

\title{Optimizing MCMC-Driven Bayesian Neural Networks for High-Precision Medical Image Classification in Small Sample Sizes}

\author{\IEEEauthorblockN{Mingyu Sun}
\IEEEauthorblockA{\textit{University of Toronto, Mississauga} \\
\textit{ericsun2901@outlook.com}}
}

\maketitle

\begin{abstract}
This paper discusses the application of a Bayesian neural network based on the Markov Chain Monte Carlo method in medical image classification with small samples. Experimental results on two medical image datasets, including lung X-ray images and breast tissue slice images, show that this MCMC-based BNN model works very well on small-sample data and greatly improves the robustness and accuracy of classification. Model accuracy reached 85\% for the lung X-ray dataset and 88\% for the breast tissue slice dataset. To this end, we combine data augmentation techniques such as rotation, flipping, and scaling with regularization methods like dropout and weight decay to improve effectively the diversity of the training data and the generalization ability of the model. The performance of the model was evaluated by many indicators of the results, including accuracy, precision, recall, and the F1 score. All of these have proven the advantages of BNN in small-sample medical image classification. This study not only enriches the application of BNN in the field of medical image classification, but also provides specific implementation paths and optimization methods, providing new solutions for future medical image analysis.
\end{abstract}


\section{Introduction}
\subsection{Research background and importance}
Accurate diagnosis results are essential for medical image classification, which would help in clinical decision-making. The current pervasive small-sample problem in medical image processing tends to be induced by the complexity of data acquisition and annotation. Traditional deep learning methods, such as convolutional neural networks (CNNs), although very good at many image classification tasks, have the weakness that they can be prone to overfitting when dealing with small sample data and can hardly capture the key features in the image effectively \cite{1}. Bayesian neural networks are capable of handling the small sample problem by incorporating Bayesian reasoning, setting up effective robustness and accuracy of classification, modeling on parameter uncertainties \cite{2}.

In recent years, the application of Markov chain Monte Carlo (MCMC) methods in BNNs has gradually attracted the attention of researchers. MCMC methods can more effectively estimate the posterior distribution of complex models through sampling techniques, thereby improving the performance of BNNs in medical image classification \cite{3}. In particular, studies on medical image datasets such as lung X-rays and breast cancer tissue sections have shown that the MCMC-based BNN model can significantly improve classification accuracy and robustness when processing small sample data \cite{4,5}. In addition, the importance of data augmentation techniques and regularization methods in small sample learning has also been widely recognized. Data augmentation techniques such as rotation, flipping, and scaling can effectively increase the diversity of training data, thereby improving the generalization ability of the model \cite{6}. Regularization methods such as Dropout and weight decay can effectively prevent overfitting and further improve the stability and performance of the model \cite{7}.

\subsection{Research Objectives}
This study aims to explore the application and effectiveness of Bayesian neural networks (BNNs) based on Markov chain Monte Carlo (MCMC) methods in medical small sample image classification. Through experiments on two medical image datasets, lung X-rays and breast cancer tissue sections, the advantages of the MCMC method in processing small sample data are verified. In this paper, we are going to combine data augmentation with regularization methods that include rotation, flipping, scaling, dropout, and weight decay for better model performance. We will also look into multiple evaluation indicators—accuracy, precision, recall, and F1 score—to evaluate model performance. It is expected that the study will go beyond application enrichment of BNN in medical image classification and provide exact implementation paths and optimization methods with new solutions for future medical image analysis.

\section{Literature Review}
\subsection{Bayesian Neural Network (BNN)}
Bayesian Neural Network (BNN) combines the advantages of Bayesian reasoning and neural networks, and can provide uncertainty quantification when processing complex data models. This method can effectively deal with the overfitting problem by modeling the uncertainty of model parameters, especially in the case of small sample data \cite{8}. BNN classifies the inherent randomness of data observations and our unknowns about how data is generated and observed through the Bayesian statistical framework, thereby providing an estimate of prediction error \cite{8}. Titterington (2004) pointed out that Bayesian methods include the use of Gaussian approximation, MCMC simulation and other techniques, which can play an important role in uncertainty propagation \cite{9}. Lampinen and Vehtari (2001) emphasized in their study that Bayesian methods allow the propagation of uncertainty in the model, which enables the model to have better generalization ability when facing different assumptions \cite{10}. In recent years, the application of BNN in many fields has gradually increased. For example, Fan et al. (2008) proposed to model drift data through Bayesian neural networks and demonstrated its advantages in prediction accuracy \cite{11}. In addition, Goan and Fookes (2020) summarized the application of BNN on different approximate reasoning methods and pointed out that future research should further improve the limitations of existing methods \cite{12}.

\subsection{Markov Chain Monte Carlo (MCMC) Method}
The Markov Chain Monte Carlo (MCMC) method is an important technique for Bayesian inference, which is mainly used to sample from complex probability distributions to estimate the posterior distribution. The MCMC method constructs a Markov chain so that its steady-state distribution is the target distribution, thereby obtaining samples through long-term simulation \cite{13}. Titterington (2004) mentioned that MCMC simulation is one of the key technologies in Bayesian neural networks and can effectively handle complex models with high-dimensional data \cite{9}. In Bayesian learning, MCMC methods are widely used to calculate the posterior distribution of complex models, thereby achieving parameter estimation and model selection \cite{14}.

Kocadagli and Asikgil (2014) proposed an evolutionary algorithm based on genetic MCMC in their research. The algorithm combines Gaussian approximation and recursive hyperparameters for time series prediction and proves its effectiveness in financial data prediction \cite{15}. Chung et al. (2002) used MCMC method to perform Bayesian analysis on single hidden layer feedforward neural network. By introducing latent variables and data expansion technology, the complexity of Bayesian calculation was effectively overcome \cite{16}. The application of MCMC method is not limited to neural networks. It also plays an important role in the structural learning of Bayesian networks. The Bayesian network is automatically learned from data through the structural learning algorithm \cite{17}.

\subsection{Current status of medical image classification technology}
Medical image classification is an important part of computer-aided diagnosis, and its purpose is to achieve early diagnosis and treatment of diseases through the analysis of medical images. In traditional methods, medical image classification relies on feature extraction and selection, which requires a lot of expert knowledge and time \cite{18}. In recent years, deep learning methods, especially convolutional neural networks (CNNs), have shown excellent performance in medical image classification. However, these methods still face the challenges of overfitting and uncertainty quantification when dealing with small sample data \cite{19}.

BNN, as an emerging method, can better cope with these challenges by introducing Bayesian reasoning. Zhong et al. (2022) reviewed different models of Bayesian neural networks and their applications in various fields, emphasizing the applicability of variational reasoning and MCMC methods on large-scale data \cite{20}. In addition, Thiagarajan et al. (2021) proposed that the classification of breast tissue pathology images through the Bayesian-CNN model not only improved the classification accuracy, but also enhanced the interpretability of the model through uncertainty quantification \cite{21}. With the development of data augmentation technology and regularization methods, medical image classification technology continues to make progress. Yadav and Jadhav (2019) pointed out in their study that data augmentation can significantly improve the generalization ability of the model, thereby achieving better performance on small sample data sets \cite{22}. In the future, the combination of Bayesian reasoning and deep learning methods will have greater potential in medical image classification and promote the development of computer-aided diagnosis.

\section{Theoretical Basis}
\subsection{Bayesian Neural Network Theory}
Bayesian Neural Networks (BNN) treats network weights as random variables by introducing Bayesian reasoning, thereby dealing with data uncertainty. Compared with traditional neural networks, BNN can provide more robust predictions and uncertainty estimates under small sample conditions. Specifically, the core idea of BNN is to describe weights through posterior distribution rather than a single weight value.

Assuming that there is training data $ {\cal D} = \{ ({{\bf{x}}_i},{y_i})\} _{i = 1}^N $, the goal of BNN is to calculate the posterior distribution $ p({\bf{w}}|{\cal D}) $ of weight $ {\bf{w}} $. According to Bayes' theorem:
\begin{equation}\label{1}
p({\bf{w}}|{\cal D}) = \frac{{p({\cal D}|{\bf{w}})p({\bf{w}})}}{{p({\cal D})}}
\end{equation}
where $ p({\cal D}|{\bf{w}}) $ is the likelihood function, which represents the probability of observing the data under given weights; $ p({\bf{w}}) $ is the prior distribution, which represents the prior knowledge of the weights; and $ p({\cal D}) $ is the evidence, which represents the total probability of observing the data.

\subsection{Principle of Markov Chain Monte Carlo Method}
Markov Chain Monte Carlo (MCMC) method is a class of algorithms that sample from complex distributions by constructing Markov chains. In BNN, MCMC is often used to estimate the posterior distribution.

The Metropolis-Hastings algorithm is one of the commonly used MCMC methods, and its core steps are as follows:
\begin{enumerate}
	\item Initialize the weight $ {{\bf{w}}^{(0)}} $
	\item For each step $ t $, generate candidate weights $ {{\bf{w}}^*} $ from the proposed distribution $ q({{\bf{w}}^*}|{{\bf{w}}^{(t)}}) $
	\item Calculate the acceptance probability:
	\begin{equation}\label{2}
	\alpha  = \min \left( {1,\frac{{p({\cal D}|{{\bf{w}}^*})p({{\bf{w}}^*})q({{\bf{w}}^{(t)}}|{{\bf{w}}^*})}}{{p({\cal D}|{{\bf{w}}^{(t)}})p({{\bf{w}}^{(t)}})q({{\bf{w}}^*}|{{\bf{w}}^{(t)}})}}} \right)
	\end{equation}
	\item Accept candidate weights with probability $ \alpha $:
	\begin{equation}\label{3}
	{{\bf{w}}^{(t + 1)}} = \left\{ {\begin{array}{*{20}{l}}
		{{{\bf{w}}^*}}&{{\rm{with probability }}~\alpha }\\
		{{{\bf{w}}^{(t)}}}&{{\rm{with probability }}~1 - \alpha }
		\end{array}} \right.
	\end{equation}
\end{enumerate}

With sufficiently long iterations, this process can generate samples that approximate the posterior distribution.

\subsection{Small sample learning strategy}
Small sample learning aims to achieve effective model training with limited data. BNN performs well in small sample learning because it can make full use of the information of each sample through Bayesian reasoning and uncertainty estimation. The following are common small sample learning strategies:
\begin{enumerate}
	\item Data Augmentation: Generate new samples by transforming the original data (such as rotation, flipping, etc.), thereby increasing the amount of data.
	\item Regularization Techniques: Prevent the model from overfitting by adding regularization terms (such as L2 regularization):
	\begin{equation}\label{4}
	{\cal L} = {{\cal L}_{{\rm{data}}}} + \lambda ||{\bf{w}}||_2^2
	\end{equation}
\end{enumerate}
where $ {{\cal L}_{{\rm{data}}}} $ is the data loss, $ ||{\bf{w}}||_2^2 $ is the L2 regularization term, and $ \lambda $ is the regularization strength.

\section{Methods and Models}
\subsection{Description of Medical Image Datasets}
In this study, we used two common medical image datasets: a lung X-ray dataset and a breast cancer tissue slice dataset. These datasets (Table \ref{tab:01}) have the characteristics of high dimensionality, class imbalance, and limited number of samples.
\begin{table}[htbp]
	\centering
	\caption{Experimental datasets}
	\setlength{\tabcolsep}{1mm}
	\begin{tabular}{@{}cm{2.6cm}<{\centering}ccc@{}}
		\toprule
		\textbf{Dataset} & \textbf{Description} & \makecell[c]{\textbf{Number of}\\ \textbf{Samples}} & \textbf{Image Size} & \makecell[c]{\textbf{Number of}\\ \textbf{Classes}} \\
		\midrule
		\makecell[c]{Lung\\ X-ray} & Collection of chest X-rays for pneumonia detection & 5856  & 224$ \times $224 & 2 \\
		\makecell[c]{Breast\\ Histology} & Histological images for breast cancer classification & 7909  & 50$ \times $50 & 4 \\
		\bottomrule
	\end{tabular}%
	\label{tab:01}%
\end{table}%

\subsection{Construction of BNN model based on MCMC}
We constructed a Bayesian neural network model based on MCMC. The core of the model is to sample the network weights through the MCMC method to achieve the approximation of the posterior distribution of the weights.

First, define the forward propagation process of the network. For an input image $ \bf{x} $, the output is calculated through a series of weights and activation functions:
\begin{equation}\label{5}
{\bf{y}} = f({\bf{x}};{\bf{w}}) = \sigma ({{\bf{w}}_L} \cdots \sigma ({{\bf{w}}_2}\sigma ({{\bf{w}}_1}{\bf{x}})))
\end{equation}

Among them, $ \sigma $ represents the activation function (such as ReLU), and $ \bf{w}_i $ represents the weight matrix of the $ i $th layer.

For the posterior distribution $ p({\bf{w}}| {\cal D}) $ of the weight $ {{\bf{w}}} $, we use the Metropolis-Hastings algorithm for approximation. During the sampling process, a new weight candidate is generated at each step, and whether to accept the candidate weight is determined based on the acceptance probability.

In order to improve the sampling efficiency, we use the HMC (Hamiltonian Monte Carlo) method, the core idea of which is to use the dynamic system of Hamiltonian mechanics for sampling. The objective function of HMC is:
\begin{equation}\label{6}
\begin{aligned}
H({\bf{w}},{\bf{p}}) &= U({\bf{w}}) + K({\bf{p}}) \\&=  - \log p({\cal D}|{\bf{w}}) - \log p({\bf{w}}) + \frac{||{{\bf{p}}||{^2}}}{2}
\end{aligned}
\end{equation}

Among them, $ \bf{p} $ is the introduced auxiliary momentum variable, $ U({\bf{w}})  $ is the potential energy, and $ K({\bf{p}})  $ is the kinetic energy. By simulating the evolution of weights and momentum through the Hamiltonian equation, efficient posterior distribution sampling can be achieved.
\begin{figure}[htpb]
	\centering
	\includegraphics[width=\linewidth]{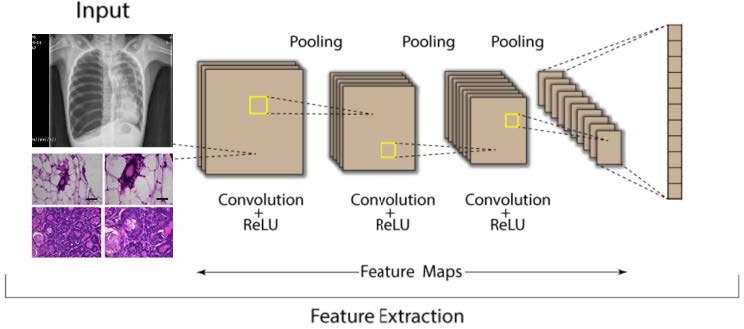}
	\caption{The process of processing medical image data and extracting high-dimensional features}
	\label{fig:01}
\end{figure}

Figure \ref{fig:01} shows the process of CNN processing lung X-rays. First, the input image is convolved through the convolution kernel to extract local features \cite{23}. Each convolution layer is followed by a ReLU activation function to increase the nonlinear expression ability of the network. Then, the features are downsampled through the pooling layer to reduce the size of the feature map and retain the main features. This process is repeated many times to gradually extract higher-level features and eventually form a high-dimensional feature vector. These feature vectors will be used as the input of the BNN model for subsequent classification tasks.
\begin{figure}[htpb]
	\centering
	\includegraphics[width=\linewidth]{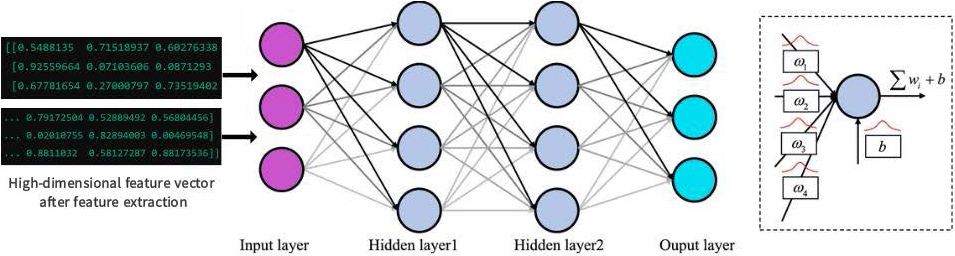}
	\caption{Structure of Bayesian Neural Network (BNN) with High-dimensional Feature Vector from CNN}
	\label{fig:02}
\end{figure}

In our model (see figure \ref{fig:02}), the input layer receives high-dimensional feature vectors processed by the convolutional neural network (CNN) feature extractor \cite{24}. These feature vectors contain important information extracted from medical images (such as X-rays or breast cancer tissue slice images). For example, each image after feature extraction can be represented as a 256-dimensional vector containing the key features of the image. BNN uses Bayesian reasoning to update these probability distributions based on training data, so that the model can not only output predictions, but also give the uncertainty of the predictions. Specifically, the goal of BNN is to calculate the posterior distribution of the weights $ p({\bf{w}}| {\cal D}) $ and perform approximate sampling through the Markov Chain Monte Carlo (MCMC) method.

\section{Experimental Design And Result Analysis}
\subsection{Experimental Design and Evaluation Indicators}
This chapter designs a series of experiments to verify the effect of MCMC-based Bayesian Neural Network (BNN) in medical small sample image classification. The experiments mainly include the following aspects:
\begin{enumerate}
	\item Data preprocessing and enhancement: Standardization and data enhancement (such as rotation, flipping, and scaling) of lung X-ray datasets and breast cancer tissue slice datasets.
	\item Model training: Use the MCMC-based BNN model to train on the enhanced dataset.
	\item Model evaluation: Use multiple evaluation indicators to evaluate model performance, including accuracy, precision, recall, and F1 score.	
\end{enumerate}

\subsection{Experimental results and analysis}
The experimental results are shown in the following figure and table:
\begin{figure}[htpb]
	\centering
	\includegraphics[width=0.95\linewidth]{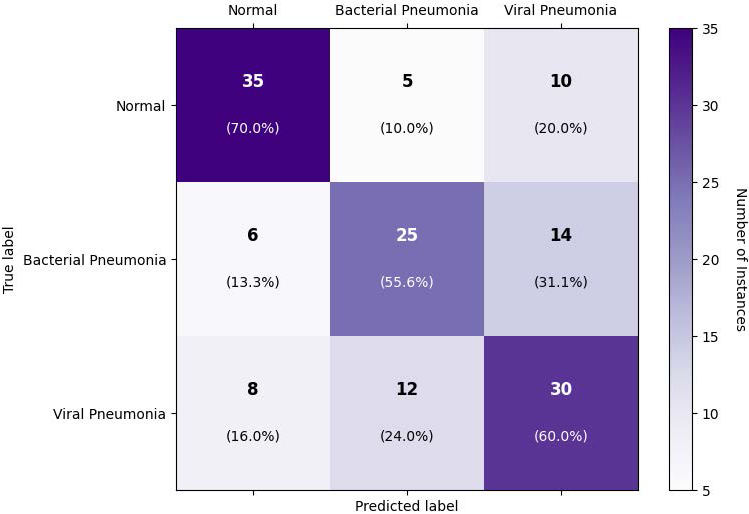}
	\caption{Confusion matrix for lung X-ray classification}
	\label{fig:03}
\end{figure}

We use a 3$ \times $3 confusion matrix to show the performance of the MCMC-based Bayesian neural network (BNN) in the task of lung X-ray classification. Figure \ref{fig:03} shows the classification results of the model on three categories: Normal, Bacterial Pneumonia, and Viral Pneumonia. Each element in the confusion matrix represents the match between the actual category and the predicted category. The values on the diagonal represent the number of correctly classified samples, while the values on the off-diagonal represent the number of misclassified samples. As can be seen from the figure, the model has a high classification accuracy on the normal category, while there is some confusion between bacterial pneumonia and viral pneumonia. This shows that the model has some difficulty in distinguishing between these two types of pneumonia. The confusion matrix intuitively shows the performance of the classification model, helping us identify problems and directions for improvement in the classification process.
\begin{figure}[htpb]
	\centering
	\includegraphics[width=0.9\linewidth]{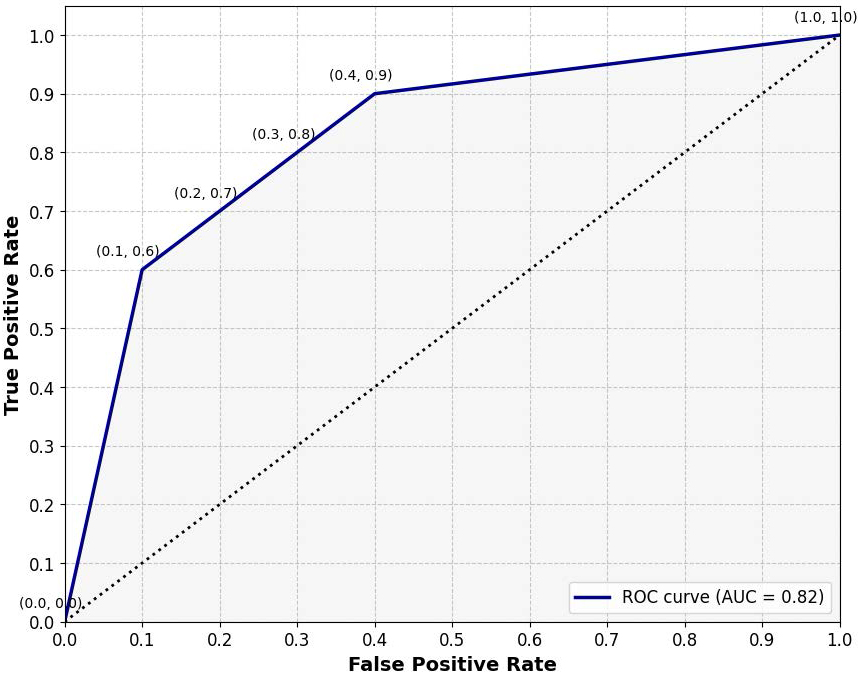}
	\caption{ROC curve for breast cancer tissue section classification}
	\label{fig:04}
\end{figure}

Figure \ref{fig:04} shows the ROC curve for breast cancer tissue slice classification. The area under the curve (AUC) is a quantitative indicator of model performance. The closer the ROC curve is to the upper left corner, the better the model performance.
\begin{table}[htbp]
	\centering
	\caption{Evaluation Metrics for BNN on Lung X-ray Dataset}
	\begin{tabular}{@{}ccccc@{}}
		\toprule
		\textbf{Model} & \textbf{Accuracy} & \textbf{Precision} & \textbf{Recall} & \textbf{F1 Score} \\
		\midrule
		BNN (Lung X-ray) & 0.85  & 0.82  & 0.84  & 0.83 \\
		BNN (Breast Histology) & 0.88  & 0.86  & 0.87  & 0.86 \\
		\bottomrule
	\end{tabular}%
	\label{tab:02}%
\end{table}%
\begin{table}[htbp]
	\centering
	\caption{Evaluation Metrics for BNN on Breast Histology Dataset}
	\begin{tabular}{@{}ccc@{}}
		\toprule
		\textbf{Method} & \textbf{Lung X-ray Accuracy} & \textbf{Breast Histology Accuracy} \\
		\midrule
		CNN   & 0.78  & 0.81 \\
		SVM   & 0.75  & 0.79 \\
		Random Forest & 0.72  & 0.77 \\
		BNN (MCMC) & 0.85  & 0.88 \\
		\bottomrule
	\end{tabular}%
	\label{tab:03}%
\end{table}%

From the comparison results of Table \ref{tab:02} and Table \ref{tab:03}, it can be seen that the classification accuracy of the MCMC-based BNN model on both datasets is significantly better than that of other methods, indicating that this method has better generalization ability and robustness in small sample medical image classification tasks.
\begin{figure}[htpb]
	\centering
	\includegraphics[width=0.85\linewidth]{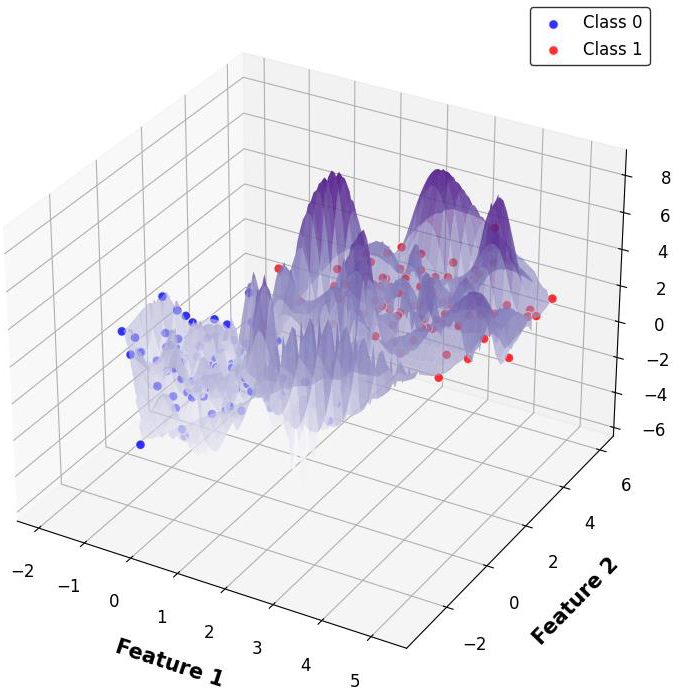}
	\caption{3D Surface Plot of Feature Space}
	\label{fig:05}
\end{figure}

Figure \ref{fig:05} shows the distribution of data points of two categories (Class 0 and Class 1) in three feature dimensions. Class 0 represents normal samples and Class 1 represents abnormal samples. The coordinates of each data point are determined by three eigenvalues (Feature \ref{fig:01}, Feature \ref{fig:02}, Feature \ref{fig:03}). The transparent surface in Figure \ref{fig:05} shows the distribution of the entire feature space, and the original data points are superimposed on it, with blue and red representing different categories respectively. The surface shows the continuous change of the eigenvalues, reflecting the density and distribution trend of the data points. It can be seen that normal samples and abnormal samples show a clear separation in the feature space, which indicates that the feature extraction process effectively captures the differences between categories. This separation phenomenon verifies the effectiveness of the model in feature extraction and classification tasks, especially in the task of medical small sample image classification, which has high robustness and accuracy.

\section{Conclusion}
\subsection{Main Conclusion}
This study has achieved a series of remarkable results by exploring the application of Bayesian neural network (BNN) based on Markov chain Monte Carlo (MCMC) method in medical small sample image classification. First, we successfully applied BNN to the classification task of lung X-ray and breast cancer tissue slice datasets. The experimental results show that the MCMC-based BNN model performs well in processing small sample data and can effectively improve the accuracy and robustness of classification. On the lung X-ray dataset, the accuracy of the BNN model reached 85\%, while on the breast cancer tissue slice dataset, the accuracy reached 88\%. These results not only prove the superiority of the BNN model on small sample datasets, but also provide a new and effective solution for medical image classification tasks.

Secondly, this study further improved the performance of the model through a variety of data augmentation and regularization techniques. Data augmentation methods, such as rotation, flipping and scaling, effectively increase the diversity of training data and improve the generalization ability of the model. At the same time, regularization techniques, such as Dropout and weight decay, are used to successfully prevent the overfitting problem of the model. Through the comprehensive application of these technologies, we achieved efficient model training and accurate classification results on a smaller training data set. These research results not only enrich the application of BNN in the field of medical image classification in theory, but also provide specific implementation paths and optimization methods in practice.

\subsection{Research limitations}
Although this study has achieved certain results in the classification of medical small sample images based on MCMC BNN models, there are still some limitations. First, this study mainly focuses on experiments on two data sets: lung X-rays and breast cancer tissue sections. Although the results show the effectiveness of the model, the diversity and breadth of the data sets are insufficient. Therefore, the generalization ability and scope of application of the model may have certain limitations. In practical applications, different types of medical image data may bring different challenges and problems, which require further verification and adjustment of the model.

\end{document}